\documentclass[12pt]{iopart}
\usepackage{iopams}

\usepackage{graphicx}
\usepackage{color}
\usepackage{pict2e}

\newcommand{\bea}{\begin{eqnarray}}
\newcommand{\eea}{\end{eqnarray}}
\newcommand{\beq}{\begin{equation}}
\newcommand{\eeq}{\end{equation}}

\newcommand{\eqref}[1]{(\ref{#1})}

\newcommand{\IZ}{{\mbox{\tiny IZ}}}
\newcommand{\NZ}{{\mbox{\tiny NZ}}}

\newcommand{\vect}[1]{\boldsymbol{#1}}

\begin{document}

\title{Spinning, Precessing,  
 Black Hole Binary Spacetime via Asymptotic Matching}

\author{
Hiroyuki~Nakano$^{1,2}$,
Brennan~Ireland$^{2}$,
Manuela~Campanelli$^{2}$
and
Eric~J.~West$^{3}$}

\address{$^1$Department of Physics, Kyoto University,
Kyoto 606-8502, Japan.}

\address{$^2$Center for Computational Relativity and Gravitation,
and School of Mathematical Sciences, Rochester Institute of
Technology, Rochester, New York 14623, USA.}

\address{$^3$Department of Physics and Astronomy,
University of Minnesota Duluth,
Duluth, Minnesota 55812, USA.}

\begin{abstract}

We briefly discuss a method to construct a global, analytic, approximate
spacetime for precessing, spinning binary black holes.
The spacetime construction is broken into three parts:
the inner zones are the spacetimes close to each black hole,
and are approximated by perturbed Kerr solutions;
the near zone is far from the two black holes,
and described by the post-Newtonian metric;
and finally the wave (far) zone, where retardation effects need to be taken
into account, is well modeled by 
the post-Minkowskian metric. These individual spacetimes are then
stitched together using asymptotic matching techniques
to obtain a global solution that approximately satisfies
the Einstein field equations.
Precession effects are introduced by rotating the black hole spin direction
according to the precessing equations of motion, 
in a way that is consistent with the global spacetime construction.

\end{abstract}

\maketitle

\section{Introduction}

In our previous paper~\cite{Ireland:2015cjj},
we constructed a global, analytic, approximate
spacetime for non-precessing, spinning binary black holes (BBHs)
in the (quasi-circular) inspiral phase.
To obtain this metric, 
we used the framework discussed
in~\cite{Yunes:2005nn, Yunes:2006iw, JohnsonMcDaniel:2009dq}.
Using this method, the spacetime is divided into various zones,
and described by appropriate approximation for each zone.
Each zone's metric is smoothly matched via the techniques of
matched-asymptotic-expansions,
and the use of transition functions (see~\cite{Yunes:2006mx})
in the overlapping regions of validity, called the \textit{buffer zones} (BZs).

The spacetime close to each black hole (BH) is called the \textit{inner zone} (IZ);
it is derived by satisfying the Teukolsky equations~\cite{Teukolsky:1973ha}, 
which describe linearized perturbations around a background Kerr BH.
There are two IZs,
IZ1 around BH1 (with mass $m_1$ and dimensionless spin parameter $\vect{\chi}_1$)
and IZ2 around BH2 (with $m_2$ and $\vect{\chi}_2$).
The post-Newtonian (PN) approach accurately describes
the weak gravitational field around two BHs, called the \textit{near zone} (NZ)
(see, e.g., \cite{Blanchet:2013haa}).
The PN expansion is a Taylor expansion in $v/c \ll 1$ (slow motion)
or $GM/(rc^2) \ll 1$ (weak fields), 
where $v$ is the characteristic velocity of the BH,
$M$ is the mass of the BH, and $r$ is the radial distance coordinate from the BH.
The region far away from the BHs (much larger than a gravitational wavelength)
is the \textit{far zone} (FZ)
and also referred to as the \textit{wave zone}.
This zone is modeled by the post-Minkowskian formalism
(see, e.g.,~\cite{Will:1996zj, Blanchet:2013haa}).

Using the above metrics and matched asymptotic expansion, 
previous work has given the global metrics on a particular spatial hypersurface,
i.e., initial data for non-spinning and non-precessing, spinning BBHs,
in~\cite{JohnsonMcDaniel:2009dq} and \cite{Gallouin:2012kb}, respectively.
In~\cite{Mundim:2013vca}, the above works were extended to be able to
describe the dynamical spacetime valid for arbitrary times.
The method was generalized further to include aligned and counter-aligned 
spins in~\cite{Ireland:2015cjj}.
Here, we extend this construction for precessing, spinning BBHs.

This paper is organized as follows.
In Section~\ref{sec:match}, we derive the coordinate transformation
between the IZ and NZ metrics in the matched asymptotic expansion.
Since the calculation is almost parallel to the derivation
discussed in~\cite{Ireland:2015cjj}, we do not repeat the detailed analysis
in the main text, and briefly summarize it in~\ref{app:details}. 
In Section~\ref{sec:PNEOM}, we discuss the PN equations of motion (EOM) briefly
as another important ingredient for the dynamical spacetime construction.
Section~\ref{sec:discussion} is devoted to discussions. 
We use the geometric unit system, where $c=G=1$.
Greek and latin letters are used as spacetime and spatial indices, respectively.

\section{Asymptotic Matching}\label{sec:match}

To describe the NZ spacetime, we restrict to the currently available explicit PN 
expressions published in the literature
for the non-spinning and spin-orbit terms in the PN harmonic (PNH) 
coordinate system ($x^{\alpha}=\{t,\,x,\,y,\,z\}$) \cite{Blanchet:2013haa}.
Hence, we use \cite{Blanchet:1998vx} for the spin independent terms up to 2.5PN order,
and~\cite{Tagoshi:2000zg, Faye:2006gx} for the non-vanishing spin terms up to 1.5PN order,
and~\cite{Bohe:2012mr} for the next-to-leading-order spin terms.
Higher order spin coupling 
terms have been calculated for the EOM
and the equations of the precession of the spins~\cite{Marsat:2014xea, Bohe:2015ana}, but not the bulk metric.
These terms will be added to the metric
in the future as they become explicitly available.

The FZ metric is presented in~\cite{Will:1996zj, Pati:2002ux, JohnsonMcDaniel:2009dq},
and asymptotically matched to the NZ in the NZ-FZ BZ
automatically~\cite{Gallouin:2012kb}.
Therefore, we focus only on the matching calculation
between the IZ and NZ metrics here.

\subsection{Coordinate rotation}

For simplicity and clarity of presentation, we discuss only IZ1 around BH1.
The treatment of BH2 is handled by changing the labels, $1 \leftrightarrow 2$.
The IZ spacetime is described by the Kerr background metric and
its perturbation~\cite{Yunes:2005ve}
under the ingoing radiation gauge condition in Cook-Scheel harmonic 
coordinates ($X^{\alpha}=\{T,\,X,\,Y,\,Z\}$)~\cite{Cook:1997qc}.

The important point here is that the Kerr spin parameter $a_1 = M_1 \chi_1$
is given in a preferred coordinate system,
where the spin is aligned along the $Z$ coordinate direction.
On the other hand, for precessing BBHs,
we need to treat an arbitrary time-dependent spin direction.
To set the spin direction along the $Z$ direction,
we consider the following rotation of the spin direction,
\begin{eqnarray}
\vect{\chi}_1 = \chi_1 (\sin \Theta_1 \cos \Phi_1, 
\sin \Theta_1 \sin \Phi_1, \cos \Theta_1) \,,
\end{eqnarray}
where $\Theta_1$ and $\Phi_1$ are functions of time.
The above spin $\vect{\chi}_1$ is transformed to $\vect{\chi}_1'$
along the $Z$ coordinate by two rotations;
first the rotation through angle $-\Phi_1$ about the $z$ axis
and then the rotation through angle $-\Theta_1$ about the $y$ axis as
\begin{eqnarray}
\vect{\chi}_1' = {\bf Y}(-\Theta_1){\bf Z}(-\Phi_1)\vect{\chi}_1
= \chi_1 (0,0,1) \,,
\label{eq:IZsetup}
\end{eqnarray}
where
\begin{eqnarray}
{\bf Z}(\alpha) =
\left( \begin{array}{ccc}
\cos \alpha & -\sin \alpha & 0 \\
\sin \alpha & \cos \alpha & 0 \\
0 & 0 & 1
\end{array} \right)
\,,
\quad
{\bf Y}(\beta) =
\left( \begin{array}{ccc}
\cos \beta & 0 & \sin \beta \\
0 & 1 & 0 \\
-\sin \beta & 0 & \cos \beta
\end{array} \right)
\,.
\end{eqnarray}

We will use a notation here ${\bf \hat Z}(\alpha)$
and ${\bf \hat Y}(\beta)$ in which the time-time component $=1$
and the time-space components $=0$
are added to ${\bf Z}(\alpha)$ and ${\bf Y}(\beta)$, respectively.
For conciseness,  it is helpful to define and use 
``tensor-like'' notations
for the above rotation matrices, i.e.,
${\cal R}^{\mu}{}_{\nu}({\bf \hat Y}(-\Theta_1),{\bf \hat Z}(-\Phi_1))
= {\bf \hat Y}(-\Theta_1){\bf \hat Z}(-\Phi_1)$.

\subsection{Matching calculation}

Now that we have rotated of the spin direction, we follow the procedure described
in~\cite{Ireland:2015cjj} directly to match the IZ to the NZ.
In the BZ between the IZ and NZ, 
the NZ metric $g^{\NZ}_{\alpha \beta}$ is expanded by using $m_{1} \ll r_1 \ll b$,
where $r_1$ denotes the distance from BH1 and $b$ is the orbital separation
in the PNH coordinates:
\begin{eqnarray}
g^{\NZ}_{\alpha \beta} =& (g_{\alpha \beta}^{\NZ})_{0}
+ \sqrt{ \frac{m_{2}}{b} }(g_{\alpha \beta}^{\NZ})_{1}
+ \left( \frac{m_{2}}{b} \right) (g_{\alpha \beta}^{\NZ})_{2}
+ {\cal{O}}(v^3) \,,
\end{eqnarray}
where $(~~)_{i}$ denotes the $i$th order quantity, and
\begin{eqnarray}
\fl
(g^{\NZ}_{\alpha \beta})_{0} = \eta_{\alpha \beta} \,,
\quad
(g^{\NZ}_{\alpha \beta})_{1} = 0 \,,
\nonumber \\
\fl
(g^{\NZ}_{\alpha \beta})_{2} = \biggl[ \frac{2m_{1}}{m_{2}} \frac{b}{(r_{1})_{0}}
+ 2 - \frac{2}{b} \left\{(\vect{r}_{1})_{0} \cdot (\vect{\hat b})_{0}\right\}
+ \frac{1}{b^{2}} \left\{3 [(\vect{r}_{1})_{0} \cdot (\vect{\hat b})_{0}]^{2} - [(r_{1})_{0}]^{2}\right\}
\biggr]
\Delta_{\alpha \beta} \,.
\label{eq:NZ_bz}
\end{eqnarray}
Here, $\eta_{\alpha \beta} = \mbox{diag}(-1,1,1,1)$ is the Minkowski metric,
$\Delta_{\alpha \beta} = \mbox{diag}(1,1,1,1)$,
and $(\vect{\hat b})_{0}=\vect{n}_{12}$ is the unit vector from BH2 to BH1.
We will also use a notation $\hat \beta^{\alpha}=\{0,\,\vect{n}_{12}\}$ later.
The difference from~\cite{Ireland:2015cjj} due to the precession
is in $(\vect{\hat b})_{0}$, which is in the $x$--$y$ plane
for the non-precessing case~\cite{Ireland:2015cjj},
but in this work is in an arbitrary direction.

As in the non-precessing case, we carry
the asymptotic matching up to $O[(m_2/b)^1]$.
For the NZ and IZ metrics,
the asymptotic matching is implemented
order by order with respect to $(m_{2}/b)^{1/2}$,
based on the relation between two metrics,
\bea
g_{\alpha\beta}^{\NZ} = \frac{\partial X^\gamma}{\partial x^\alpha}
\frac{\partial X^\delta}{\partial x^\beta} g_{\gamma\delta}^{\IZ} \,.
\eea
Here, we consider the matching of the IZ metric $g_{\gamma\delta}^{\IZ}$ to
the NZ metric for an arbitrary spin direction.
Since the spin terms do not enter into the matching calculation
up to $O[(m_2/b)^1]$,
except for the spin direction,
the result can be easily obtained as described below.

The equations to derive the coordinate transformation
are only slightly modified from~\cite{Ireland:2015cjj}
(see~\ref{app:details} for the detailed calculation.
In the main text, we present only 
the schematic derivation).
The coordinate transformation up to first order ($O[(m_2/b)]^{1/2}$),
\eqref{eq:upto1st}, is rewritten as
\begin{eqnarray}
(X^{\alpha})_{\{1\}} &= 
{\cal R}^{\alpha}{}_{\beta}({\bf \hat Z}(\Phi_1),{\bf \hat Y}(\Theta_1)) 
\biggl[
\tilde x^{\beta}
- \sqrt{\frac{m_2}{b}} \sqrt{\frac{m_{2}}{m}} \, \tilde y_{\rm C} \, \hat t^{\beta} \biggr] \,,
\end{eqnarray}
where we have used $\{1\}$ to describe the leading $+$ first order quantity,
the total mass $m=m_1+m_2$, and 
$\tilde x^{\beta}$, $\tilde y_{\rm C}$ and $\hat t^{\beta}$ are defined
in~\eqref{eq:1}, \eqref{eq:2} and \eqref{eq:3}, respectively.
It is noted that the expression in the bracket
of the above equation is same as (20) in~\cite{Ireland:2015cjj}.

In what follows, we derive the second order coordinate transformation
for precessing, spinning, BBHs. In doing so, we first generalize equation (24)
in~\cite{Ireland:2015cjj} to include the rotation of the spin direction due to the precession
(see~\eqref{eq:for2ndCT}). The solution to~\eqref{eq:for2ndCT} is given
by applying the rotation
${\cal R}^{\alpha}{}_{\beta}({\bf \hat Z}(\Phi_1),{\bf \hat Y}(\Theta_1))$,
to the solution derived in (28) of~\cite{Ireland:2015cjj}:
\begin{eqnarray}
(X^{\alpha})_{2} =& 
{\cal R}^{\alpha}{}_{\beta}({\bf \hat Z}(\Phi_1),{\bf \hat Y}(\Theta_1))
\eta^{\beta\gamma}
(X_{\gamma})_{2,{\rm nonP}} \,,
\label{eq:final2nd}
\end{eqnarray}
where:
\begin{eqnarray}
\fl
(X_{\alpha})_{2,{\rm nonP}} =& 
\left( 1+\frac{m_2}{2 m} \right) (\tilde{x}^{\beta} \hat t_{\beta}) \hat t_{\alpha} 
+ \left( 1-\frac{\tilde x_{\rm C}}{b} \right) \Delta_{\alpha i} \tilde{x}^{i}
+ \frac{\Delta_{ij} \tilde{x}^{i} \tilde{x}^{j}}{2b} \hat \beta_{\alpha}
+ \frac{m_2}{2m} \tilde y_{\rm C} \hat \nu_{\alpha}
\cr 
\fl
& - \frac{1}{b^{2}} \left( (r_1)_0^2 \tilde x_{\rm C} \hat \beta_{i}
 - \tilde x_{\rm C}^{2} \tilde{x}_{i} \right) \delta_{\alpha}^{i} 
+ \frac{1}{3b^{2}} \left( (r_1)_0^3
- 3 \tilde x_{\rm C}^{2} (r_1)_0 \right) \hat t_\alpha \,,
\end{eqnarray}
where $\tilde x_{\rm C}$ and $\hat \nu_{\alpha}$
are defined in~\eqref{eq:4} and \eqref{eq:5}, respectively.

The explicit expressions for the coordinate transformation
are also derived in a similar fashion:
\begin{eqnarray}
X^{\alpha} =& 
{\cal R}^{\alpha}{}_{\beta}({\bf \hat Z}(\Phi_1),{\bf \hat Y}(\Theta_1))
X^{\beta}_{\rm nonP} \,,
\end{eqnarray}
where $X^{\alpha}_{\rm nonP}$ are given as:
\begin{eqnarray}
\fl
T_{\rm nonP} &=& 
t-\sqrt {{\frac {m_2}{r_{12}}}}\sqrt {{\frac {m_2}{m}}}{\tilde y_{\rm C}}
+\frac{m_2}{{r_{12}}}\, \left( \frac{1}{3}\,{\frac {{{\tilde r_1}}^{3}-3\,{{\tilde x_{\rm C}}}^{2}{\tilde r_1}}{{r_{12}}^{2}}}
 \right) 
+ \frac{5}{384} \frac{(2m + m_2)(r_{12}^3-r_{12}(0)^3)}{m^2 m_1} \,,
\cr
\fl
X^i_{\rm nonP} &=& {\tilde x}^i
+ \frac{m_2}{{r_{12}}}\, 
\biggl( 
\left( 1-{\frac {{\tilde x_{\rm C}}}{r_{12}}} \right)\, {\tilde x}^i
+ \frac{1}{2}\,{\frac {{{\tilde r_1}}^{2}}{r_{12}}} n_{12}^i
+ \frac{1}{2}\,{\frac {m_2\,{\tilde y_{\rm C}}}{m}} \lambda_{12}^i
-{\frac {{{\tilde r_1}}^{2}{\tilde x_{\rm C}}\,n_{12}^i 
-{{\tilde x_{\rm C}}}^{2}{\tilde x}^i}{{r_{12}}^{2}}}
\biggr) \,,
\end{eqnarray}
where $T_{\rm nonP} \equiv X^0_{\rm nonP}$, and $\lambda_{12}^i$ is defined
in~\eqref{eq:5}.
Here, we have used
the evolution of the orbital separation $b = r_{12} = r_{12}(t)$,
and introduced $\tilde r_1=\sqrt{\tilde x^{i}\tilde x_{i}}(=(r_1)_0)$.
In~\cite{Ireland:2015cjj}, the orbital plane is always in the $x$--$y$ plane.
For precessing BBHs, the orbit is fully 3 dimensional.
Therefore, $X^i_{\rm nonP}$ has a vector form
rather than that specified only by the equatorial orbit.

\section{Post-Newtonian Equations of Motion}\label{sec:PNEOM}

Another important ingredient to complete the dynamical spacetime construction for our 
precessing, spinning BBHs is the introduction of the PN EOM in the
harmonic gauge. 
Up to 3.5PN order, and for maximal spin
($|\vect{\chi}_\mathrm{a}|\simeq1$), 
spin-orbit effects contribute to the EOM at 1.5PN, 2.5PN, and 3.5PN. 
Spin-spin effects contribute at 2PN and 3PN. 
Cubic-in-spin effects contribute at 2.5PN and 3.5PN. 
Quartic- and quintic-in-spin effects enter at 3PN and 3.5PN, respectively. 
Following~\cite{Will:2005sn, Blanchet:2006gy, Blanchet:2013haa}, we use the Tulczyjew spin 
supplementary condition (SSC) to define a spin vector with conserved Euclidean norm. 
For this SSC, the higher order spin-orbit terms have been derived 
in~\cite{Faye:2006gx, Blanchet:2006gy, Marsat:2012fn, Bohe:2012mr}, 
see also~\cite{Blanchet:2013haa}. The next-to-leading order spin-spin terms were derived 
in~\cite{Bohe:2015ana}, and the leading order cubic-in-spin terms were derived 
in~\cite{Marsat:2014xea}. To date, leading order quartic- and quintic-in-spin contributions 
to the EOM have not been derived in PNH coordinates with this SSC. 
The full EOM will be shown in a future paper~\cite{Westetal_in_prep},
where we present the PN EOM in a ready-to-use form.

\section{Discussion}\label{sec:discussion}

We have derived here a new global, analytic, approximate spacetime
for precessing BBHs. The construction follows closely the methods employed
in~\cite{Ireland:2015cjj}. In~\cite{Ireland:2015cjj},
we tested the validity of the global metric for non-precessing, spinning BBHs
by using the Ricci scalar to estimate the violations
to the Einstein vacuum field equations,
and the relative Kretschmann invariant to discuss a normalized violation.
The difference between \cite{Ireland:2015cjj} and this paper
is only the precession due to misaligned spins.
Since the precession time scale is much larger than the orbital one,
the violations of the new approximate spacetime
will be similar to those evaluated in~\cite{Ireland:2015cjj}.

Our expectation is that this new spacetime can be used directly in 
general relativistic magnetohydrodynamic and hydrodynamic simulations 
to study the circumbinary disk around the BBH and individual mini disks
around each BH for long time evolutions in the inspiral regime
without back reaction (see, e.g., circumbinary disk~\cite{Noble:2012xz},
mini disks~\cite{dennisminidisk}).

\ack

H.N. is supported by
MEXT Grant-in-Aid for Scientific Research on Innovative Areas,
``New Developments in Astrophysics Through Multi-Messenger Observations
of Gravitational Wave Sources'', Nos.~24103001 and 24103006,
and JSPS Grant-in-Aid for Scientific Research (C),~16K05347.
B.I. and M.C. are supported by NSF
grants AST-1516150 and AST-1028087, PHY-1607520 and PHY-1305730.

\appendix

\section{Details on the matching calculation}\label{app:details}

The matching calculation in the BZ between the IZ and NZ
is almost parallel to 
the discussion given in~\cite{Ireland:2015cjj}.

\subsection{Zeroth-order matching: $O[(m_{2}/b)^0]$}

As the zeroth order, we may consider the matching as
\begin{eqnarray}
(g^{\NZ}_{\alpha \beta})_{0} &= (A_{\alpha}{}^{\gamma})_{0} (A_{\beta}{}^{\delta})_{0} (g^{\IZ}_{\gamma \delta})_{0}
\,.
\end{eqnarray}
Here, $A_{\alpha}{}^{\beta} = \partial_{\alpha} X^{\beta}$
(where $\partial_\alpha = \partial/\partial x^\alpha$),
and $(g^{\NZ}_{\alpha \beta})_{0} = (g^{\IZ}_{\alpha \beta})_{0} = \eta_{\alpha \beta}$.
Using 
\begin{eqnarray}
{\cal R}^{\alpha}{}_{\mu}({\bf \hat Z}(\Phi_1),{\bf \hat Y}(\Theta_1))
{\cal R}^{\beta}{}_{\nu}({\bf \hat Z}(\Phi_1),{\bf \hat Y}(\Theta_1))
\eta_{\alpha\beta} = \eta_{\mu\nu} \,,
\end{eqnarray}
and taking into account the position of BH1 and the spin direction, we have
\begin{eqnarray}
(X^{\alpha})_{0} = {\cal R}^{\alpha}{}_{\beta}({\bf \hat Z}(\Phi_1),{\bf \hat Y}(\Theta_1)) \tilde x^{\beta} \,.
\label{eq:0thCT}
\end{eqnarray}
where
${\cal R}^{\alpha}{}_{\beta}({\bf \hat Z}(\Phi_1),{\bf \hat Y}(\Theta_1))$
denotes the inverse transformation of 
${\cal R}^{\alpha}{}_{\beta}({\bf \hat Y}(-\Theta_1),{\bf \hat Z}(-\Phi_1))$,
and
\begin{eqnarray}
\tilde x^{\alpha} = x^{\alpha} - \frac{m_{2}}{m} b \,\hat \beta^{\alpha} \,,
\label{eq:1}
\end{eqnarray}
where $(r_1^i)_0 = \tilde x^{i}$ in the NZ metric of~\eqref{eq:NZ_bz}.
Since $\hat \beta^{\alpha}$ has a time dependence,
the time derivative of~\eqref{eq:0thCT} becomes
\begin{eqnarray}
\partial_t (X^{\alpha})_{0} 
&=& \hat t^{\alpha} 
- \sqrt{\frac{m_2}{b}} \sqrt{\frac{m_{2}}{m}} 
\, {\cal R}^{\alpha}{}_{\beta}({\bf \hat Z}(\Phi_1),{\bf \hat Y}(\Theta_1)) \hat \nu^\beta
 \,,
\label{eq:corrT}
\end{eqnarray}
in our current analysis of the matching calculation.
Here, we have defined
\begin{eqnarray}
\hat t^\alpha=\{1,\,0,\,0,\,0\} \,,
\label{eq:3}
\end{eqnarray}
and 
\begin{eqnarray}
\hat \nu^\alpha = \{0,\,\vect{\lambda}_{12} \} = \partial_t \hat \beta^\alpha/\Omega
\,.
\label{eq:5}
\end{eqnarray}
Raising and lowering tensor indices are done by the Minkowski metric,
e.g., $\hat t_\alpha=\eta_{\alpha\beta}\hat t^\beta = \{-1,\,0,\,0,\,0\}$.
Here, the time derivative of $\Theta_1$ and $\Phi_1$
is derived from the spin-orbit coupling,
and higher order ($O[(m_{2}/b)^{3/2}]$). Therefore, we have ignored it here.

\subsection{First-order matching: $O[(m_{2}/b)^{1/2}]$}

The matching equation at the first order is written as
\begin{eqnarray}
(g^{\NZ}_{\alpha \beta})_{1} =&
(A_{\alpha}{}^{\gamma})_{0} (A_{\beta}{}^{\delta})_{0} (g^{\IZ}_{\gamma \delta})_{1}
+ 2\,(A_{(\alpha}{}^{\gamma})_{1} (A_{\beta)}{}^{\delta})_{0} (g^{\IZ}_{\gamma \delta})_{0} \,,
\end{eqnarray}
where $T_{(\alpha\beta)}=(T_{\alpha\beta}+T_{\beta\alpha})/2$ denotes a symmetric tensor,
and $(g^{\NZ}_{\alpha \beta})_{1} = (g^{\IZ}_{\alpha \beta})_{1} = 0$.
Using~\eqref{eq:corrT}, we have $\partial_i (X^t)_{0}=0$
and 
$\partial_i (X^j)_{0}={\cal R}^{j}{}_{i}({\bf \hat Z}(\Phi_1),{\bf \hat Y}(\Theta_1))$~\footnote{$(A_{\alpha}{}^{\beta})_0$ is written as
a compact form,
\bea
(A_{\alpha}{}^{\beta})_0 = \partial_{\alpha} (X^{\beta})_0
= {\cal R}^{\beta}{}_{\alpha}({\bf \hat Z}(\Phi_1),{\bf \hat Y}(\Theta_1))
+ \sqrt{\frac{m_2}{b}} \sqrt{\frac{m_{2}}{m}} 
\, \hat t_\alpha 
\, {\cal R}^{\beta}{}_{\mu}({\bf \hat Z}(\Phi_1),{\bf \hat Y}(\Theta_1)) 
\hat \nu^\mu \,.
\nonumber
\eea},
the above equation becomes
\begin{eqnarray}
\eta_{\gamma\delta} {\cal R}^{\gamma}{}_{(\alpha}({\bf \hat Z}(\Phi_1),{\bf \hat Y}(\Theta_1))
(A_{\beta)}{}^{\delta})_{1} 
+ \sqrt{\frac{m_{2}}{m}} \,\hat t_{(\alpha} \hat \nu_{\beta)} = 0 \,,
\end{eqnarray}
where the second term of the left hand side arises from
the zeroth order coordinate transformation. The solution is obtained as
\begin{eqnarray}
(X^{\alpha})_{1} &= - \sqrt{\frac{m_{2}}{m}} \tilde y_{\rm C} \, {\cal R}^{\alpha}{}_{\beta}({\bf \hat Z}(\Phi_1),{\bf \hat Y}(\Theta_1))
\hat t^{\beta}
= 
- \sqrt{\frac{m_{2}}{m}} \tilde y_{\rm C} \, \hat t^{\alpha} \,,
\end{eqnarray}
where we have defined
\begin{eqnarray}
\tilde y_{\rm C} = \hat \nu_{\alpha} \tilde x^{\alpha} \,.
\label{eq:2}
\end{eqnarray} 
Since $(X^{\alpha})_{1}$ has only the time component,
there is no effect due to the precession, i.e.,
${\cal R}^{\alpha}{}_{\beta}({\bf \hat Z}(\Phi_1),{\bf \hat Y}(\Theta_1))
\hat t^{\beta} = \hat t^{\alpha}$.

\subsection{Second-order matching: $O[(m_{2}/b)^1]$}

In a similar analysis given in~\cite{Gallouin:2012kb},
we obtain $(M)_{0}=m_1$ from the divergent part
in $(r_1)_0 = |\tilde x^{i}| \to 0$. And using the zeroth-order matching,
the tidal field $(\bar{\cal E}_{ij})_{0}$
which is the perturbation around the Kerr BH in the IZ calculation,
is derived as
$(\bar{\cal E}_{ij})_{0} = \delta_{ij} - 3 \breve \beta_i \breve \beta_j$,
where $\breve \beta_i$ denotes the spatial components of 
$\breve \beta_{\alpha} = \eta_{\alpha\beta} \breve \beta^\beta 
= \eta_{\alpha\beta} {\cal R}^{\beta}{}_{\gamma}
({\bf \hat Z}(\Phi_1),{\bf \hat Y}(\Theta_1)) \hat{\beta}^{\gamma}$.
This $(\bar{\cal E}_{ij})_{0}$ is used to evaluate
the tidal tensor in (12) of~\cite{Ireland:2015cjj}.

Next, we calculate the second order coordinate transformation.
The leading and first order matching gave
\begin{eqnarray}
(X^{\alpha})_{\{1\}} &= 
{\cal R}^{\alpha}{}_{\beta}({\bf \hat Z}(\Phi_1),{\bf \hat Y}(\Theta_1)) \tilde x^{\beta}
- \sqrt{\frac{m_2}{b}} \sqrt{\frac{m_{2}}{m}} \, \tilde y_{\rm C} \, \hat t^{\alpha} \,.
\label{eq:upto1st}
\end{eqnarray}
The formal expression for the second order matching is written as
\begin{eqnarray}
(g^{\NZ}_{\alpha \beta})_{\{2\}} =& (A_{\alpha}{}^{\gamma})_{\{2\}} 
(A_{\beta}{}^{\delta})_{\{2\}} (g^{\IZ}_{\gamma \delta})_{\{2\}} \,,
\end{eqnarray}
where $\{2\}$ denotes the leading $+$ first order $+$ second order quantity.
$(A_{\alpha}{}^{\gamma})_{\{2\}}$ includes
not-yet-determined $(X^{\gamma})_{2}$ as
\begin{eqnarray}
\fl
(A_{\alpha}{}^{\gamma})_{\{2\}} &= \partial_\alpha (X^{\gamma})_{\{2\}}
= 
{\cal R}^{\gamma}{}_{\alpha}({\bf \hat Z}(\Phi_1),{\bf \hat Y}(\Theta_1)) 
\cr
\fl 
& \quad
+ \sqrt{\frac{m_2}{b}} \biggl[ \sqrt{\frac{m_{2}}{m}} \,\hat t_{\alpha} 
{\cal R}^{\gamma}{}_{\mu}({\bf \hat Z}(\Phi_1),{\bf \hat Y}(\Theta_1)) \hat \nu^{\mu}
- \sqrt{\frac{m_{2}}{m}} \, \hat \nu_{\alpha} 
{\cal R}^{\gamma}{}_{\mu}({\bf \hat Z}(\Phi_1),{\bf \hat Y}(\Theta_1))
\hat t^{\mu} \biggr]
\cr
\fl
& \quad
+ \frac{m_2}{b} 
\biggl[ - \frac{1}{b} \left(\tilde x_{\rm C} + \frac{m_2}{m} b \right) \, 
\hat t_{\alpha} {\cal R}^{\gamma}{}_{\mu}({\bf \hat Z}(\Phi_1),{\bf \hat Y}(\Theta_1))
\hat t^{\mu} + \partial_{\alpha} (X^{\gamma})_{2} \biggr] \,,
\end{eqnarray}
where
\begin{eqnarray}
\tilde x_{\rm C} = \hat \beta_{\alpha} \tilde x^{\alpha} \,.
\label{eq:4}
\end{eqnarray}
Although the above expression is slightly more complicated than
that in~\cite{Ireland:2015cjj}, we can see the relation,
${\cal R}^{\gamma}{}_{\alpha}({\bf \hat Z}(\Phi_1),{\bf \hat Y}(\Theta_1))$
in this paper $
\leftrightarrow \delta^{\gamma}{}_{\alpha}$ in~\cite{Ireland:2015cjj}.
Finally, we may solve
\begin{eqnarray}
\fl
2{\cal R}^{\gamma}{}_{(\alpha}({\bf \hat Z}(\Phi_1),{\bf \hat Y}(\Theta_1))
&(A_{\beta) \gamma})_{2} =
\biggl[ \biggl(2-\frac{2}{b} \tilde x_{\rm C} \biggr) \Delta_{\alpha \beta}
+ \frac{2}{b} \tilde x_{\rm C} \hat t_{\alpha} \hat t_{\beta} 
+ \frac{m_2}{m} \hat t_{\alpha} \hat t_{\beta}
+ \frac{m_{2}}{m} \hat \nu_{\alpha} \hat \nu_{\beta} \biggr]
\cr
\fl &
+ \biggl[ \delta_{\alpha}^{i} \delta_{\beta}^{j} \frac{2}{b^{2}}
\Bigl( \tilde x_{\rm C}^2 \delta_{ij} - (r_1)_0^2 \hat \beta_{i} \hat \beta_{j} \Bigr) \biggr]
\nonumber \\
\fl &
+ \biggl[ ( \delta_{\alpha}^{i} \hat t_\beta + \delta_{\beta}^{i} \hat t_\alpha ) \frac{1}{3b^{2}}
\Bigl( 3 (r_1)_0 \tilde{x}_{i} 
- \frac{3}{(r_1)_{0}} \tilde x_{\rm C}^2 \tilde x_{i} - 6 (r_1)_{0} \tilde x_{\rm C} \hat \beta_{i} \Bigr) \biggr]
\,.
\label{eq:for2ndCT}
\end{eqnarray}
The above solution is given in \eqref{eq:final2nd}.

\section*{References}

\bibliographystyle{iopart-num}
\bibliography{./notes.bbl}

\end{document}